\documentclass{article}
\usepackage{spconf,amsmath,graphicx}
\usepackage{amssymb}

\usepackage{graphicx}
\usepackage{amsmath}
\usepackage[utf8]{inputenc} 
\usepackage[T1]{fontenc}
\usepackage{lmodern}
\usepackage{amsmath}
\usepackage{amssymb}
\usepackage{mathrsfs}
\usepackage{graphicx}
\usepackage{xcolor}
\usepackage{float}
\usepackage{microtype}
\usepackage{amssymb} 
\usepackage{multicol}
\usepackage{graphicx} 
\usepackage{fancyhdr} 
\usepackage{eqnarray}
\usepackage{subcaption}
\usepackage{lipsum}
\usepackage{titlesec}
\usepackage{pgfgantt}
\DeclareMathAlphabet\mathbfcal{OMS}{cmsy}{b}{n}
\usepackage{etoolbox}   
\usepackage{verbatim}   
\usepackage{color}
\usepackage{hyperref}
\newcommand{\tens}[1]{%
  \mathbin{\mathop{\otimes}\displaylimits_{1}}%
}
\definecolor{c1}{HTML}{808080} 
\definecolor{c2}{HTML}{8FED8F} 
\definecolor{c3}{HTML}{4DBA17} 
\definecolor{c4}{HTML}{ADD9E6} 
\definecolor{c5}{HTML}{FF9E00} 
\definecolor{c6}{HTML}{B3804D} 
\definecolor{c7}{HTML}{FF0080} 
\definecolor{c8}{HTML}{99994C} 
\usepackage{makecell}
\usepackage{multirow}

\title{Learnable Wavelet Packet Transform for Data-Adapted Spectrograms}
\name{Frusque Gaëtan and Fink Olga\thanks{This study was supported by the Swiss Innovation Agency (lnnosuisse) under grant number: 47231.1 IP-ENG.}}
\address{ETH Zurich, Swiss Federal Institute of Technology, Zurich, Switzerland}
\begin{document}
%
\maketitle
\begin{abstract}
Capturing high-frequency data concerning the condition of complex systems, e.g. by acoustic monitoring, has become increasingly prevalent. Such high-frequency signals typically contain time dependencies ranging over different time scales and different types of cyclic behaviors. Processing such signals requires careful feature engineering, particularly the extraction of meaningful time-frequency features. This can be time-consuming and the performance is often dependent on the choice of parameters. 
To address these limitations, we propose a deep learning framework for learnable wavelet packet transforms, enabling to learn features automatically from data and optimise them with respect to the defined objective function. The learned features can be represented as a spectrogram, containing the important time-frequency information of the dataset.
We evaluate the properties and performance of the proposed approach by evaluating its improved spectral leakage and by applying it to an anomaly detection task for acoustic monitoring.

\end{abstract}
\begin{keywords}
Wavelet Packet Transform, Sparse Decomposition, High Frequency Signals, Anomaly Detection, Harmonic Analysis.
\end{keywords}
\section{Introduction}
\label{sec:intro}


Capturing high-frequency data concerning the condition of complex systems, e.g. by acoustic monitoring, has become increasingly prevalent.  
As a first step, learning tasks based on high-frequency signals typically require an extraction of relevant time-frequency features before they are used as inputs in machine learning (ML) algorithms. ML algorithms are generally designed for inputs of a fixed size.
Fourier and wavelet transforms provide very good feature extraction capabilities and are able to reduce the dimensionality of the data. However, their application requires time-consuming engineering of the features that is dependent on both the dataset and the problem.

Learning the features automatically from raw data can help to overcome the challenges of manual feature engineering. Deep learning (DL) methods like Convolutional Neural Networks (CNN) are able to automatically learn representative features from raw signals \cite{zhang2017new}, \cite{huang2018aclnet}. However, their performance on time-series data is highly dependent on the hyperparameters  and the architecture. Both of these are typically specific to the dataset. Furthermore, 
the resulting algorithms are not easily interpreted by the domain experts. 
Recently, several architectures that combine the advantages of signal processing and the learning capabilities of neural networks have shown promising results. Some examples include successful applications on speaker recognition with kernels that are constrained to correspond to band-pass filters only \cite{ravanelli2018speaker}, sound classification with a learned filter-bank \cite{sailor2017unsupervised}, molecular partners prediction, and utilisation of a learned wavelet transform in the field of cosmology \cite{ha2021adaptive}. These models turned out to be more interpretable or to contain fewer parameters than a standard CNN.



In this work, we propose a deep learning architecture, based on wavelet packet transform (WPT), called L-WPT. Our approach provides an automatically learnt sparse time-frequency representation of a univariate signal.
While the previously proposed learnable wavelet transform approaches \cite{xiong2020novel}, \cite{recoskie2018learning}, \cite{jawali2019learning}, \cite{ha2021adaptive} focus on learning the best filter to use throughout the entire architecture, in this research, we seek to generalise the filter learning and add learned biases for automatic denoising. While a similar approach was recently proposed in \cite{michau2021fully} in the context of discrete wavelet transforms (DWT), the advantage of the WPT structure over DWT is that it decomposes the signal in spectral bands of similar width and with a similar time resolution, enabling a better spectral resolution on the high frequencies. Consequently, L-WPT can be seen as an ideal tool to learn data-adapted spectrograms, concurrently reducing the number of parameters in the neural network significantly. 
We demonstrate that the proposed approach is capable of mitigating the spectral leakage problem that is inherent to WPT. We further show the applicability of the proposed architecture for detecting anomalies in acoustic monitoring dataset.

\section{Background}
\label{sec:Background}
\subsection{Wavelet Packet Transform (WPT)}
\label{sec:WPT}
The discrete WPT, introduced in \cite{coifman1992wavelet}, projects the signal on uniform frequency bands of desired size. The WPT has a multi-level structure and can be considered a multiresolution analysis since the output of the current level is recursively used as input to the next level. The basic bloc of a WPT applied to an input signal $\mathbf{y}$ is:
\begin{align}\label{eq:filt}
& (\mathbf{y}^{lp})_{(n)} = ( \mathbf{h}^{lp} * \mathbf{y} )_{(2n)}, \hspace{0.3cm}  (\mathbf{y}^{hp})_{(n)} = ( \mathbf{h}^{hp} * \mathbf{y} )_{(2n)}. 
\end{align}
where $*$ is the convolution operation and $\mathbf{y}^{lp} $ ($\mathbf{y}^{hp} $) corresponds to the low- (high-) pass filtered input data with a cut-off frequency of $\pi /2$. This transformation doubles the frequency resolution (the frequency content of each wavelet coefficient spans half the input data frequency) to the detriment of a halved time resolution ($\mathbf{y}^{lp} $ and $\mathbf{y}^{hp} $ each contains half the number of samples in $\mathbf{y}$). Recursively, we use $\mathbf{y}^{lp} $ and $\mathbf{y}^{hp} $ as input of the same bloc eq.~(\ref{eq:filt}) to generate frequency bands of desired width. This process gives as many outputs (also called nodes) as $2^L$, where $L$ is the number of recursions (called layers).


\subsection{Perfect reconstruction}
\label{sec:kWPT}
It is possible to recover the input signal $\mathbf{y}$ from $\mathbf{y}^{lp} $ and $\mathbf{y}^{hp} $ via inverse WPT (iWPT). The following constraints on the filters are sufficient to ensure a perfect reconstruction \cite{strang1996wavelets}; we separate them into relational constraints between filters and constraint on the filter's coefficients.\\
$\bullet$ \textbf{Architectural properties}: $\mathbf{h}^{hp}$ has to be the Alternating Flip (AF) of $\mathbf{h}^{lp}$. It causes both filters to have mirrored magnitude responses around $\pi /2$. The AF of a filter $\mathbf{h}$ is denoted as ${\rm flip}[\mathbf{h}]$ and corresponds, for a filter of size $K+1$ with $K$ odd and $ k \in \{0,...,K\}$, to:
\begin{align}\label{eq:filt2}
 {\rm flip}[\mathbf{h}]_{(k)} = (-1)^k (\mathbf{h})_{(K-k)}.
\end{align}
Also, the iWPT filters have to respect the Anti-Cancellation Conditions (ACC) to remove aliasing terms due to the use of non-ideal filters. This can be done by considering the delayed paraconjugate of $\mathbf{h}^{hp}$ and $\mathbf{h}^{lp}$. Applied to a real filter $\mathbf{h}$, this operation results in $\mathbf{\bar{h}} $ with:
\begin{align}\label{eq:filt3}
& (\mathbf{\bar{h}})_{(k)} = (\mathbf{h})_{(K-k)}.
\end{align}
Notice that by respecting both AF and ACC properties, we only have to design one filter $\mathbf{h}^{lp}$.\\
$\bullet$ \textbf{Kernel property}: to ensure the conservation of the frequency content, $\mathbf{h}^{hp}$ and $\mathbf{h}^{lp}$ have to be power complementary (the power sum of both filters has to be one for all frequencies). Considering $\mathbf{h}^{hp}$ is the AF of $\mathbf{h}^{lp}$, then, these filters are conjugate mirrors \cite{mallat1999wavelet}, which implies that $\mathbf{h}^{lp}$ coefficients have to be normalised and double-shift orthogonal (for non-null shifts). 


\subsection{Denoising with the WPT}
\label{sec:dWPT}
Signal denoising is one of the major applications of wavelet analysis \cite{bayer2019iterative}. It has been shown that, concerning regular and structured signals, a wavelet transform will lead to a sparse decomposition \cite{mallat1999wavelet}. We can then assume that the noise will correspond to wavelet coefficients of small amplitudes.
To eliminate these low-amplitude wavelet coefficients, we consider the sharp, hard-thresholding activation function proposed in \cite{michau2020feature} and denoted as ${\rm HT}[x]$:
\begin{align}\label{eq:act}
{\rm HT}[x; \gamma]=x ( \sigma_{ (-10(x+ \gamma))} +  \sigma_{(10(x- \gamma))}   ),
\end{align}
with $ \sigma_{(x)} = 1/(1+e^{-x})$  the sigmoid function and $\gamma$ the bias acting as thresholds on both sides of the origin.

\section{Method: Learned-WPT}
\label{sec:L-Background}

The proposed method is a learnable deep architecture based on WPT and referred to as Learned-WPT (L-WPT). This section details the proposed methodology.

\subsection{Architecture of L-WPT}
L-WPT resembles the WPT structure. However, the filter elements for each bloc and each layer are learned independently. The activation function eq.~(\ref{eq:act}) is applied to the output of each filter. We impose the filters to respect all the architectural properties presented in section~\ref{sec:kWPT}. This has the advantage of reducing the number of filters that need to be learnt by a factor of four, to keep the mirrored frequency interpretation and the ACC property. 
However, the kernel property can't be preserved, which will prevent the perfect reconstruction of the input signal.

By applying L-WPT, the wavelet coefficients at layer $l$, node $i$ (denoted $\mathbf{y}_l^{i}$) are convoluted with the filter $\theta^{i}_{l}$ to produce the wavelet coefficients of the upper layer: 
\begin{align}\label{eq:filtL}
 (\mathbf{y}_{l+1}^{2i} )_{(n)} &  ={\rm HT}[ ( \mathbf{\theta}_{l}^{i}  * \mathbf{y}_{l}^{i} )_{(2n)} ;  \gamma_{l+1}^{2i}], \\
(\mathbf{y}_{l+1}^{2i+1} )_{(n)} & ={\rm HT}[ ( {\rm flip}[ \mathbf{\theta}_{l}^{i}]  * \mathbf{y}_{l}^{i} )_{(2n)} ;  \gamma_{l+1}^{2i+1}],
\end{align}
It is simply a rewriting of eq.~(\ref{eq:filt}), considering a different filter for each bloc, the AF  eq.~(\ref{eq:filt2}) and the learned activation function eq.~(\ref{eq:act}). 
Similar to the iWPT, it is possible to estimate the wavelet coefficients $\mathbf{y}^{i}_l $ from the upper layer via inverse L-WPT.  The computation of the estimated reconstruction of $\mathbf{y}^{i}_l$, considering eq.~(\ref{eq:filt3}) to respect the ACC property, is denoted $\hat{\mathbf{y}}^{i}_l $ and is written:
\begin{align}
\mathbf{\hat{y}}^{i}_l =  \mathbf{\overline{\theta^{i}_{\textnormal{l}}}} *   {\rm up} \hspace{-0.1cm} \left[\hat{\mathbf{y}}^{2i}_{l+1}\right] + \mathbf{\overline{{\rm flip}[\theta^{i}_{\textnormal{l}} ]}} *   {\rm up} \hspace{-0.1cm} \left[\hat{\mathbf{y}}^{2i+1}_{l+1}\right] 
\end{align}
The upsampling operation ${\rm up} [\bullet ]$ is necessary to counteract the effect of the downsampling in eq.~(\ref{eq:filtL}); it can be defined as ${\rm up} [\mathbf{x} ]_{(2n)}=x_{(n)}$ and  ${\rm up} [\mathbf{x} ]_{(2n+1)}=0$. 

\begin{figure}[h]
\centering
\includegraphics[width=\columnwidth]{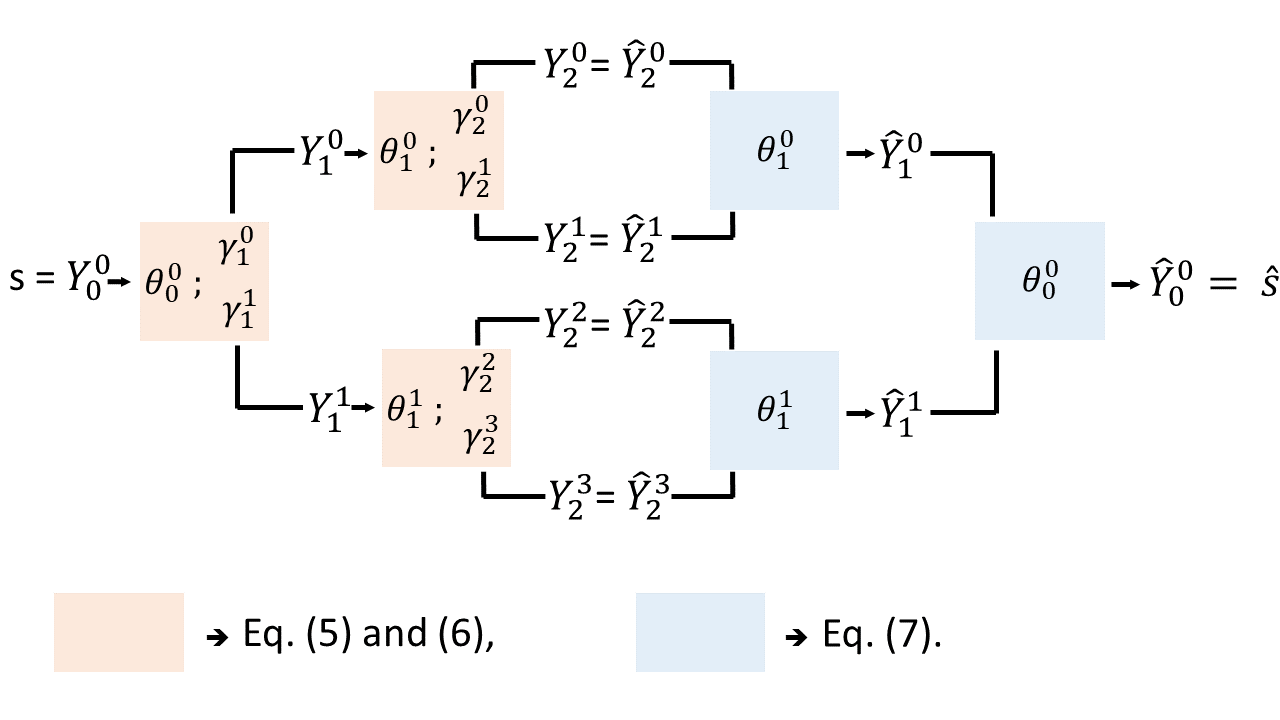}
\caption{Decomposition tree of L-WPT with $L=2$}\label{fig:1}
\end{figure}

Fig.~\ref{fig:1} illustrates the decomposition and reconstruction via L-WPT showing the decomposition tree for the case of $L=2$ layers. 

\subsection{Loss for data-adapted spectrogram}
\label{sec:OL-WPT}
We refer to ${\mathbf{Y}_{L} = [\mathbf{y}^{0}_{L}, \mathbf{y}^{1}_{L},..., \mathbf{y}^{2^L-1}_{L}]}$ as the set of all node outputs for the last layer $L$. The matrix $|\mathbf{Y}_{L}|$ can be considered a learned spectrogram representing the temporal evolution of $2^L$ frequency bands. 
We propose to impose sparsity on the obtained spectrogram. We aim to learn the best representation of the signal  with the filters ${ \mathbf{\Theta}=[\mathbf{\theta}_0^0, \mathbf{\theta}_1^0, \mathbf{\theta}_1^1,...,  \mathbf{\theta}_{L-1}^{2^{L-1}-1}]}$ as high-amplitude wavelet coefficients that are sparse. The activation function eq.~(\ref{eq:act}) with learned bias ${\mathbf{\Gamma}   = [ \gamma^{0}_{1}, \gamma^{1}_{1},...,  \gamma^{2^{L}-1}_{L} ]}$ will suppress the low wavelet coefficient associated with noise and induce sparsity in $|\mathbf{Y}_L|$. Inspired by \cite{michau2020feature}, the proposed loss function to be minimised is defined as:
\begin{align}\label{Final_func}
\underset{ \mathbf{\Theta}, \mathbf{\Gamma}    }{\rm argmin} \hspace{0.5cm}   \mid\mid \mathbf{s} -  \hat{\mathbf{s}}  \mid\mid_1 + \alpha  \mid\mid \mathbf{Y}_{L}   \mid\mid_1, 
\end{align}
where $\mathbf{s}=\mathbf{y}_0^0$ is the input data and $\hat{\mathbf{s}}=\hat{\mathbf{y}}_0^0$ is the reconstructed signal.
$\mid\mid \bullet   \mid\mid_1$ is the $\ell_{1}$-norm function and corresponds to a surrogate of the non-convex $\ell_{0}$-norm to induce sparsity. $\alpha$ is the trade-off parameter between the sparsity of the spectrogram and the quality of the reconstruction.  The $\ell_{1}$-norm is used for both the residual and the regularisation to ease the selection of the trade-off parameter. Indeed, by choosing $\alpha=1$, we give the same importance to both parts of the eq.~(\ref{Final_func}) since $\mathbf{Y}_{L}$ and $\mathbf{s} $ contain the same number of elements.  


\subsection{Learning strategy}
The proposed framework can be implemented as a deep auto-encoding architecture. The encoding blocs correspond to two CNN layers (one per filter) with a shared kernel, stride 2 for the downsampling and followed by the activation functions eq.~(\ref{eq:act}). The decoding blocs contain the corresponding transposed CNN layers with upsampling.
Considering $L$ layers and $K+1$ coefficients per filter, the number of trainable parameters is $\sum_{l=1}^{L} {2^{(L-l)}K}  $ for the filters and $\sum_{l=1}^{L} {2^{L}} $ for the bias.

\section{Evaluation case studies}
\label{sec:copyright}
For both applications, we use WPT and L-WPT with $L=5$ layers, which involves an embedding $\mathbf{Y}_{L} $ of 32 nodes (now denoted $\mathbf{Y} $ since we no longer consider the previous layers). We initialise the filters of L-WPT using the Daubechies wavelets with 8 coefficients (denoted db4), and the biases are initialised to 0.3. The L-WPT is trained using the Adam optimiser with a learning rate of $0.001$.
\subsection{Spectral leakage analysis}
\begin{figure}[h]
\begin{subfigure}[b]{0.45\columnwidth}
        \centering \includegraphics[width=\textwidth]{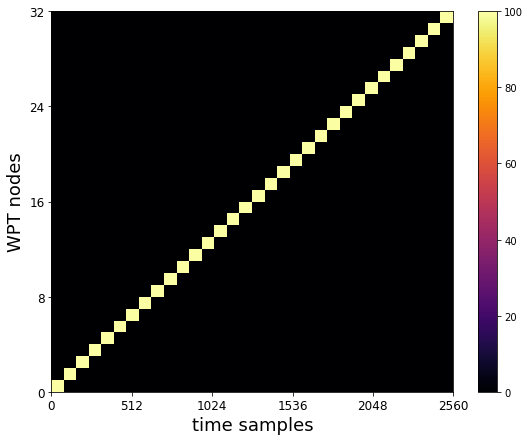}
        \caption{Ideal}
    \end{subfigure}
    \begin{subfigure}[b]{0.45\columnwidth}
            \centering \includegraphics[width=\textwidth]{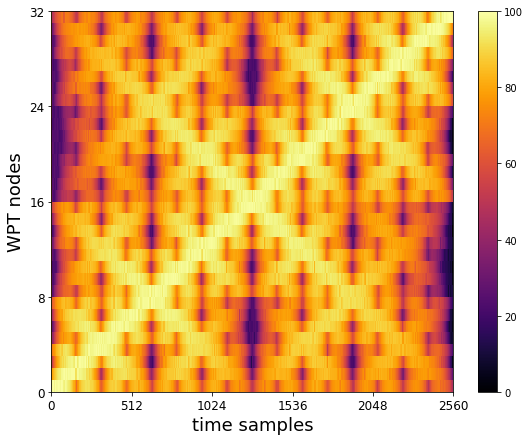}
            \caption{WPT-db4}
        \end{subfigure}

\begin{subfigure}[b]{0.45\columnwidth}
        \centering \includegraphics[width=\textwidth]{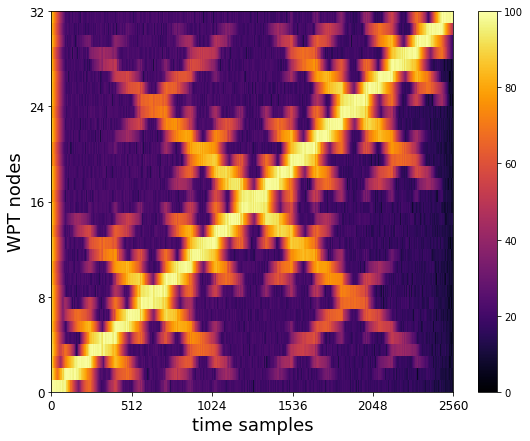}
        \caption{WPT-butt}
    \end{subfigure}
    \begin{subfigure}[b]{0.45\columnwidth}
            \centering \includegraphics[width=\textwidth]{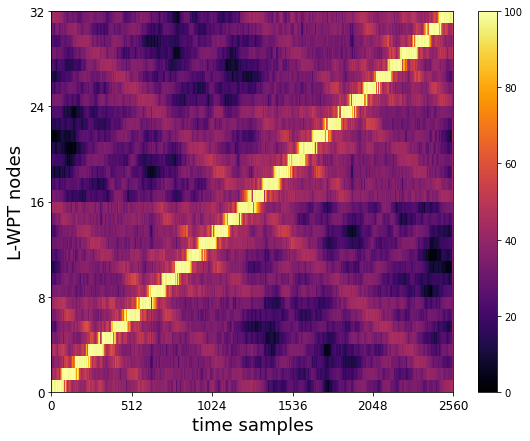}
            \caption{L-WPT}
        \end{subfigure}  
             
\caption{Output of the pure FSC $\mathbf{w}_{\beta=0} $ through WPT with different filters (a), (b), (c), and our L-WPT (d). The colormap is based on the \% of the maximum value. }\label{fig:WPT}

\end{figure}

Using non-ideal filters in WPT generates a spectral leakage and makes  the discrimination of harmonic content imperfect. To evaluate the performance of our framework, we propose, based on \cite{bruna2016selection}, to measure the level of distortion produced by L-WPT. For this experiment, we generate a 10-seconds Frequency-Swept Cosine (FSC), with a sampling frequency of $f_s=2^{13}$, which sweeps from null to the maximum frequency $f_s/2$ such as:
\begin{align}
w_{\beta}(n) = {\rm cos}( \pi  0.05 n^2  ) + \beta \mathcal{N}_{(0,1)} ,
\end{align}
where $n \in \{ 1,...,10f_s\}$, $\mathcal{N}_{(0,1)} $ is a realisation of a normal distribution and $\beta$ is the noise level.
The ideal WPT (based on ideal filters) of the pure FSC ($\mathbf{w}_{\beta=0} $) consists of each node successively activating each of the $80$ samples according to the sweep level (see fig.~\ref{fig:WPT}(a)). To assess the quality of the transform, we propose computing the Root Sum Square (RSS) of the difference between the estimated representation and the ideal one. 
Both representations are first normalised to have the same l2-norm.


To learn the coefficients and biases of L-WPT, we generate 400,000 noisy cosinus signals with a random frequency. Each can be written as ${s(n) ={\rm cos}(2 \pi f n) + \mathcal{N}_{(0,1)}}$, where $f$ is a realisation of a uniform distribution taking values from $0$ to $f_s/2$.

The time-frequency representation of the signal $\mathbf{w} $ using L-WPT is compared to the results of WPT with different wavelet families or filters. We consider the standard Daubechies wavelet with 8 and 46 coefficients (WPT-db4 and WPT-db23). We also consider the best filters for measuring the fluctuating harmonics using WPT, which are, according to \cite{bruna2016selection}: Butterworth with 12 coefficients (WPT-butt) and Elliptic with 8 coefficients (WPT-ellip). In contrast to the Daubechies wavelet, these are Infinite Impulse Response (IIR) filters, which should improve the frequency selectivity. For the three last filters, we have selected the number of coefficients (in the range from 2 to 76) which best minimise the RSS score.

\begin{figure}[h]
\centering \includegraphics[width=\columnwidth]{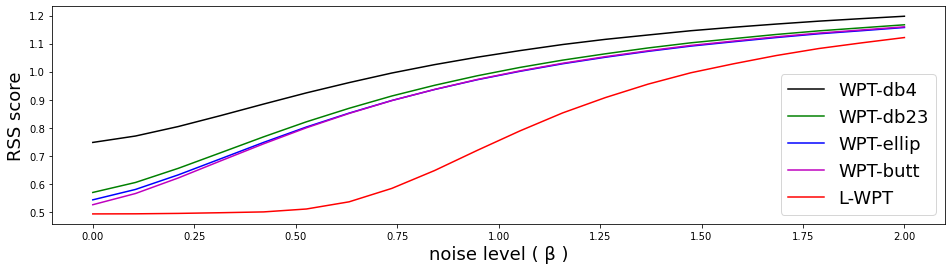}    
\caption{RSS score for WPT with different filters and L-WPT in function of the noise level $\beta$  }\label{fig:WPT2}
\end{figure}

Figs.~\ref{fig:WPT} (b), (c), and (d) show the spectral leakage for WPT-db4, WPT-ellip, and L-WPT in decibels (DB). We can observe the consequent spectral leakage introduced by the commonly used  WPT-db4. The spectral discrimination is improved with WPT-butt. However, it seems to be impacted by border effects and delays. Finally, the spectrogram obtained with L-WPT is close to the ideal one.
Fig.~\ref{fig:WPT2} shows the mean RSS score over 100 realisations of $\mathbf{w}_{\beta}$ for $\beta \in \{0,0.1,0.2,...2\}$. For all noise levels, L-WPT outperforms all other methods. For pure FSC ($\beta=0$), the RSS score for the L-WPT is 0.49 and 0.53 for the WPT-butt (second-best method). Our method is robust to noise. The biggest gap is obtained at $\beta=0.6$, where the RSS score for L-WPT is 0.54 and 0.85 for WPT-butt.



\subsection{Anomaly detection with L-WPT}
We evaluate the performance of our model on an anomaly detection task with the Sound Dataset for Malfunctioning Industrial Machine Investigation and Inspection (MIMII). It consists of audio recordings of four types of industrial machines, \textit{i.e.}, valves, pumps, fans, and slide rails. 
The training and validation dataset is set as in \cite{purohit2019mimii}. The training dataset contains sound recordings from healthy machines only. 
We aim to detect anomalies without any knowledge of the fault characteristic. 

After learning a data-adapted spectrogram representation of the healthy sounds via L-WTP, a 1-class Extreme Learning Machine (ELM) with 100 neurons \cite{michau2020feature} is applied for fault discrimination with the input as the mean and maximum of the residual and of each frequency band (following \cite{michau2021fully}).

We compare the embedding learned by the L-WPT with a) the standard WPT-db4, b) the learned DWT with $L=17$ layers (called DesPawn) \cite{michau2020feature}, and c) the architecture of the WPT-CNN from \cite{xiong2020novel} (contrary to the L-WPT, it does not have an activation function and the same learned filter is used throughout the entire structure). For each data-driven method, the results may vary with the number of epochs and the hyperparameter $\alpha$. We fix $\alpha=1$ for the L-WPT and DesPawn and provide the best area under the curve (AUC) over 500 epochs resulting from the classification of the 1-class ELM. 

Table~\ref{Tab:t2} shows the AUC scores for each machine and each method. DesPawn is outperformed by the WPT methods (except for the fan case). This demonstrates the importance of having a better spectral resolution at high frequencies for this application. The L-WPT gives the best results overall, outperforming the WPT by 1.9\% and the learned WPT-CNN by 1.2\%. It outperforms other approaches on each machine except for the valve, where the WPT-CNN yields a slightly better result. Here, the level of sparsity $\alpha$ seems to be too high.

\begin{table}
\centering
\begin{tabular}{|c|c|c|c|c||c|}
\hline 
Machine &  Fan &Pump &Slider & Valve  & All  \\
\hline
\hline
WPT & 91.3 & 85.6 & 97.8 & 97.0 &   92.9   \\
\hline
DesPawn & 92.4 & 82.2 & 92.5 & 94.0  &  90.3   \\
\hline
WPT-CNN &  91.8 & 86.8 &  98.5 & \textbf{97.3} &  93.6 \\
\hline
L-WPT & \textbf{94.9} &\textbf{87.7} &  \textbf{99.5} &  97.0 &  \textbf{94.8 }\\
\hline
\end{tabular} 
\caption{mean AUC for the different machine types}\label{Tab:t2}
\end{table}

\section{Conclusion}
We propose a new deep architecture that integrates the learning abilities of deep learning with the advantages of signal processing approaches. The proposed L-WPT is able to learn a data-adapted spectrogram of the signal. It achieves superior performance in discerning pure harmonic content and outperforms other similar methods on an anomaly detection task for acoustic monitoring. 

There are several ways to extend the proposed framework, such as integrating additional constraints into the objective function, analysing the interpretability of the proposed framework, and extending it to multivariate signals.



\vfill\pagebreak

\bibliographystyle{ieeebib}
\bibliography{strings,refs}

\end{document}